\documentclass[%
 reprint,
 amsmath,amssymb,
 aps,prl,
floatfix
]{revtex4-2}
\usepackage{xcolor}
\usepackage{graphicx}
\usepackage{dcolumn}
\usepackage{bm}
\begin{document}

\title{A sheet-like structure in the proximity of compact DNA}

\author{Garima Mishra}
\email{garima.mishra@ashoka.edu.in}
\author{Somendra M. Bhattacharjee}
\email{somendra.bhattacharjee@ashoka.edu.in}
\affiliation{Department of Physics,  Ashoka University, Sonepat 131029, India}%

\date{\today}

\begin{abstract}
  We determine the phase diagram of DNA with inter- and intra-strand
  native-pair interactions that mimic the compaction of DNA.  We
  show that DNA takes an overall sheet-like structure in the region
  where an incipient transition to a compact phase would have
  occurred.  The stability of this phase is due to the extra entropy
  from the folding of the sheet, which is absent in the remaining
  polymer-like states of the phase diagram.
\end{abstract}

\maketitle

\section{Introduction}

A common bacterium like a rod-shaped {\it E.coli} of size $2\mu$m$\times
1\mu$m$\times1\mu$m stores a $1.5$mm long DNA of $N=4\times 10^6$ base
pairs.  Even if the DNA takes an entropy-driven random coil form its
size $\sim \sqrt{N}\sim 5\mu$m would still be too large to fit in the
cell \footnote{The persistence length of DNA is $\xi=50$ nm or
  $150$bp.  The radius of gyration for a random walk of $N=4\times
  10^6/150$ units each of length $\xi$ is $R\sim
  \sqrt{N}\xi/\sqrt{6}\sim 5\mu$m.  One may take the Kuhn units of
  length $2\xi$, but that does not affect the order of magnitude
  estimate.  If the excluded volume interactions are taken into
  account the size of the swollen conformation would be much larger.
  If excursions beyond the average size is taken into account, the
  actual size would be larger than the estimate above.  }.  The
evolutionary strategy of encapsulating protein-compactified
chromosomes in a nucleus encased by membranes, as in eukaryotes, does
not apply to prokaryotes that lack a nucleus, but still, compaction
would resolve the DNA-packing puzzle.  Over the years, with support
from the dynamic light scattering \cite{dias} and the high-resolution
single-molecule imaging \cite{stracy} studies, the proposal of a
boundaryless region called nucleoid \cite{joyeux,verma} is gaining
ground, which creates an environment for DNA to collapse into a
compact form.

DNA in a salt solution behaves as a polymer in a good solvent, called
the swollen or coil phase, where excluded volume interactions dominate
the overall behaviour \cite{flory,Dietler}.  A surprising result was
that a long DNA, much larger than its persistence length, can be
collapsed into a compact shape \cite{bloom1997} by adding ethanol,
different types of cations, hydrophilic polymers, or organic
solvents \cite{teifreview,ke2010,mikhailenko}.  The collapse of DNA
remained elusive in earlier studies because of the requirement of
large lengths, though the tendency of liquid-crystalline
ordering \cite{livolant} was recorded for oligonuceotides. This
tendency of local arrangement persists in the collapsed phase \cite{bloom1997}, 
suggesting the condensed DNA phase \cite{note2} 
to be different from the globular (collapsed) phase of a polymer 
below the theta point \cite{flory}.  Moreover, as all base pairs should be easily
accessible, an ordered or correlated structure would have a functional
advantage.  Consequently, a fractal globule phase has been proposed as
a possible phase, where DNA is compact but without any
knots \cite{nechaev}---a phase different from the usual compact
globule. Such a fractal phase with power-law correlations of monomers
finds support in the human genome \cite{chromo}.

As a tertiary structure of DNA, the compact phase is found to be
insensitive to the secondary structure like B, A, or Z \cite{condBA}.
DNA also undergoes a temperature (or pH) induced melting transition
and a force-induced unzipping transition, both involving cooperative
breakings of the hydrogen bonds of the base pairs \cite{smb99}.  In a
bacterial environment, melting or unzipping is required for
replication and, at least locally, for transcription to make the bases
accessible.  With the DNA in the nucleoid, a question
arises: does a compact DNA melt?  Thus the phenomenon of DNA
compaction in nucleoids is a topic of interest in biology, polymer
physics, and also for practical applications because of the ease of
insertion of a compact DNA in bio-systems, e.g., in gene
therapy \cite{hansma}.

The compaction phenomenon has been studied in some detail
by treating DNA as polymers or polyelectrolytes. 
For example,  a recent study focussed on the melting temperature 
variations in the presence of attraction exclusively between bound 
pairs \cite{debj}. 
Other proposals include compaction induced by phase separation,
mechanisms based on neutralizing charges on DNA, or the presence of small 
cations in solutions \cite{joyeux}. In sec II, we introduce the model and the details of the simulation.
The various phases and the phase diagram are determined in sec III where a heuristic argument is also presented to motivate the phases.
The summary of the findings is in sec IV. Some of the details are given  in the supplementary material in the form of figures.

\section{Model and Method} 
\begin{figure}[t]
\includegraphics[scale=0.33,trim={0.1in 1.8in 5.8in 1.5in},clip] {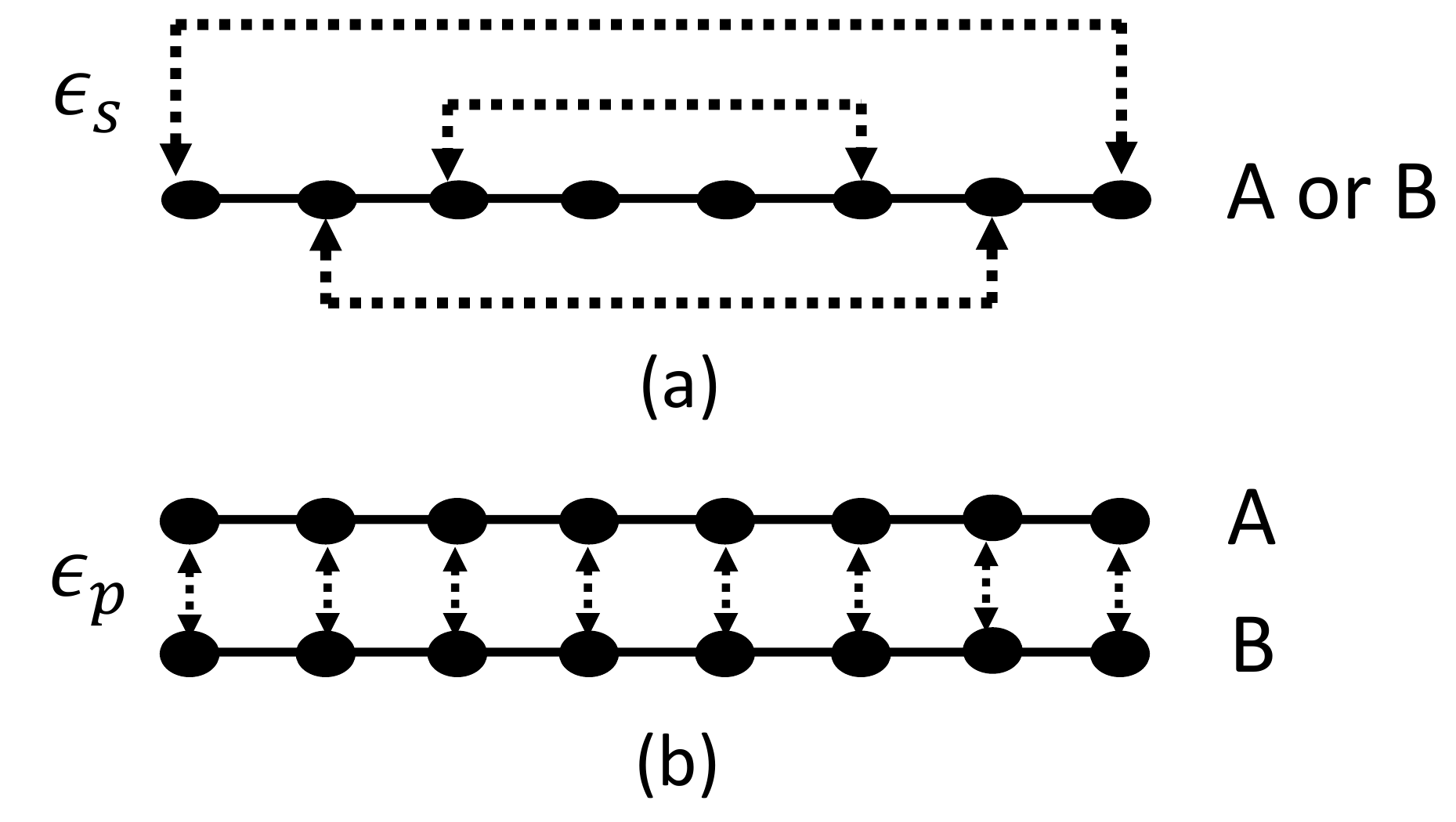}
\caption{Schematic diagram for the native contacts: (a) intra-strand
  contacts (dotted lines)in each polymer chain (A or B) with energy
  $\epsilon_s$ and (b) base-pair contacts (dotted lines) between
  chains A and B with energy $\epsilon_p$.}
 \label{fig:schematic}
\end{figure}

We explore how compaction modifies double-stranded DNA (dsDNA)
melting. The interactions are chosen to favour a folded conformation
of dsDNA, thereby avoiding a generic globular phase in a poor solvent.
Our coarse-grained model consists of two single-stranded DNA (ssDNA),
A and B (also called polymers), with native DNA base pairing and an
additional intra-strand base-pairing type interaction that allows each
polymer to fold on itself.  The model potential considered for polymer
A or B is
\begin{eqnarray}
 \left. \begin{array}{ll}
 E_A/k_BT\\ 
 E_B/k_BT
 \end{array}\right\}& = &  \sum_{i=1}^{N-1}K(d_{i,i+1}-d_0)^2 + 
 4 \sum_{\rm N-nat}\bigg(\frac{\sigma_{i,j}}{d_{i,j}}\bigg)^{12}\nonumber\\
 & &+4{\epsilon_{s}}\sum_{\rm N.C.} \bigg[\bigg(\frac{\sigma_{i,j}}{d_{i,j}}\bigg)^{12}  -\bigg(\frac{\sigma_{i,j}}{d_{i,j}}\bigg)^6\bigg],\label{eq:1}
\end{eqnarray}
where N.C. and N-nat denote native and non-native contacts.  Here,
each polymer consists of $N (=36)$ beads, the distance between beads
$d_{i,j}$ is defined as $|\bm{ r}_i-\bm{r}_j|$, where $\bm{r}_i$ and
$\bm{r}_j$ are the position vectors of beads $i$ and $j$, respectively
with $i,j\in [1,N]$.  We use dimensionless distances with
$\sigma_{i,j}=1$ and $d_0=1.12$. The energy parameters in the
Hamiltonian are in units of $k_B T$ where $k_B$ is the Boltzmann
constant and $T$ is the temperature.  The harmonic (first) term on the
rhs with dimensionless spring constant $K (=100)$ couples the adjacent
beads along the chain \cite{MishraJCP,MishraPRL}.  The second term is
a repulsive potential that prevents overlap of non-native pairs of
monomers of chain A (and B) \cite{MishraPRL,Allen,Smit}.  The third
term is the van der Waals energy (involving $\epsilon_s$) that allows
pairing of non bonded monomers at position $i$ and $N-i+1$(Fig.
\ref{fig:schematic}a).  These intra-strand pairings are the native
contacts (N.C.) in Eq.  (\ref{eq:1}).  The inter-strand interactions
are given by
\begin{eqnarray}
 \frac{E_{A,B}}{k_BT} & = &  \sum_{\rm N. C.}4{\epsilon_{p}}\bigg[\bigg(\frac{\sigma_{i,j}}{d_{i,j}}\bigg)^{12} -
\bigg(\frac{\sigma_{i,j}}{d_{i,j}}\bigg)^{6}\bigg] \nonumber \\
& & + \sum_{\rm N-nat}4\bigg(\frac{\sigma_{i,j}}{d_{i,j}}\bigg)^{12}.\label{eq:2}
\end{eqnarray}
The base pairing between chains A and B is considered using the first
term on the rhs of Eq. \ref{eq:2}. The native base-pair contacts (same
$i$ of both the chains, see Fig. \ref{fig:schematic}b) are such that
it results in a ladder structure of dsDNA \cite{MishraJCP, MishraPRL}.
The second repulsive term of the potential energy in Eq. \eqref{eq:2}
prevents non-native pairings of monomers of chains A and B
\cite{MishraJCP,MishraPRL,Allen,Smit}. 
We obtain the dynamics of system by solving the set of Brownian equations
\cite{kumar2010biomolecules}

\begin{equation}
\frac{d\bm{r_i}}{dt} = \frac{1}{\zeta}(\bm{F}_c + \bm{\Gamma}).
\label{Brow}
\end{equation}

Here,  $\bm{F}_c=-\bm{\nabla} E$ is the conservative force and $E$ is the sum of $E_A$, $E_B$ and $E_{A,B}$. $\zeta$ is the friction coefficient, and $\bm{\Gamma}$ is  the random force, 
a white noise with zero mean and correlation
$\langle\Gamma_i(t)\Gamma_j(t')\rangle=2\zeta k_B T\delta_{i,j}\delta(t-t')$.
 The temperature of the simulation is set to be constant $T=0.16$. The equation of motion is integrated using a Euler method. We obtained the phase diagram from our
simulation by monitoring the peak position in energy fluctuation as a
function of intra-strand energy ($\epsilon_s$) at a fixed base-pairing
energy ($\epsilon_p$) and vice versa. 

\section{Results}

\subsection{Phase diagram in $\epsilon_s-\epsilon_p$ plane}

To understand the phase behaviour, let us consider a simpler model
with energy based on contact numbers.  The interaction energy can be
written as
\begin{equation}
  \label{eq:1p}
  \frac{E}{k_BT}= -\epsilon_p n_{\rm{AB}}  - \epsilon_s (n_{\rm{AA}}+n_{\rm{BB}}),
\end{equation}
where $n_{\rm{AB}}$ is the number of native base pairs between the two
chains, with $-\epsilon_p$ as the energy for each pair, $n_{\rm{AA}}$,
$n_{\rm{BB}}$ are the number of intra-strand pairs with $-\epsilon_s$
as the energy per intra-strand pair. For a given DNA, $\epsilon_s$,
$\epsilon_p$ change with temperature with $\epsilon_s/\epsilon_p=$
constant.

In the $\epsilon_s$-$\epsilon_p$ plane, there are four fixed points
(FP) or limiting points which are representatives of the four phases
one would expect naively, viz,
\begin{itemize}
\item State-I:  a  denatured phase of two free ssDNA chains (FP: $\epsilon_s=\epsilon_p=0$)
\item State-II:  bound A-B as dsDNA in a  swollen phase (FP: $\epsilon_s=0, \epsilon_p=\infty$)
\item State-III: two folded ssDNA in the denatured phase (FP: $\epsilon_s=\infty, \epsilon_p=0$) 
\item State-IV: dsDNA in the folded state (FP: $\epsilon_s=\epsilon_p=\infty$)
\end{itemize}
Assuming an all-or-none state, i.e., either all pairs are formed or
all broken, the free energy of phase I is $F_I/k_BT= -S_I,$ while for
phase II, it is $F_{II}/k_BT=-N \epsilon_p - S_{II},$ where $S_{I}$
and $S_{II}$ are the total entropies of the chains in the respective
phases.  A first-order transition then takes place at
$\epsilon_p=(S_I-S_{II})/N$, and the phase boundary is expected to be
vertical in the $\epsilon_s$-$\epsilon_p$ plane. By symmetry, the
transition from I to III should also be at the same value of
$\epsilon_s$, and the boundary is horizontal, independent of
$\epsilon_p$. Similar vertical or horizontal boundaries are expected
for the III$\leftrightarrow$IV and the II$\leftrightarrow$IV
transitions.  Deviations from linearity might occur near the
intersection of the four transition lines because of the proximity of
other phases.  Can there be other intermediate phases in this 
two-parameter phase diagram?

\begin{figure}[t]
\begin{center}
 \includegraphics[scale=0.37,trim={2.5in 1.0in 2.75in 0.5in},clip] {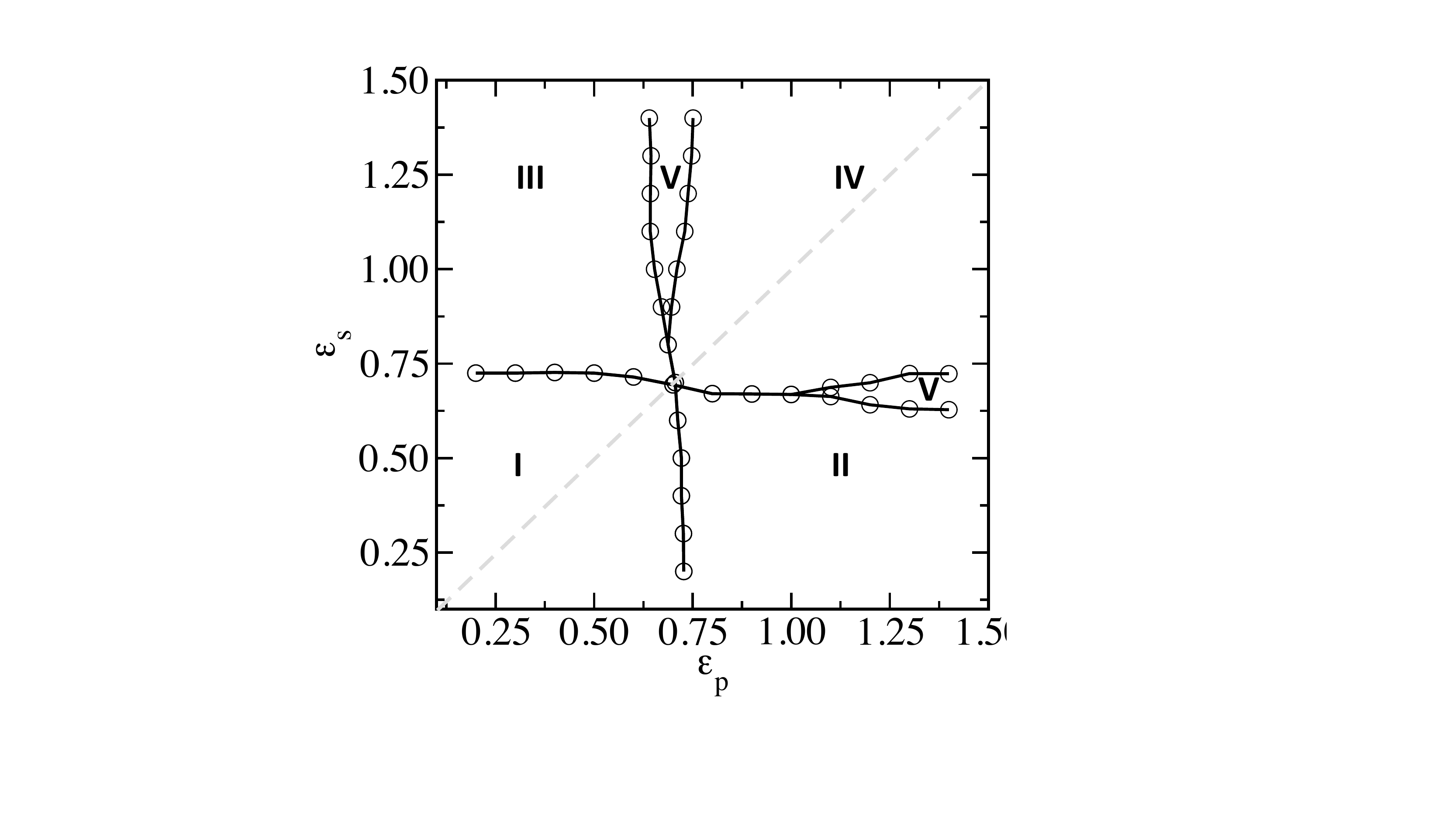}   
\end{center}
\caption{The phase-diagram shows the occurrence of various states in different
  regions of the $\epsilon_s$-$\epsilon_p$ plane.  These states are 
  State-I (two free ssDNA), State-II(dsDNA), State-III (2 folded ssDNA), State-IV 
  (folded dsDNA), and State-V (sheet). The dashed line 
  corresponds to $\epsilon_s=\epsilon_p$ and goes through the 
  intersection point of the phase boundaries.
 }
 \label{fig:phase_dia}
\end{figure}

\begin{figure*}[t]
\vspace{-0.2in}
\includegraphics[scale=0.5 ,trim={0.3in 0.5in 0.1in 1.5in},clip]{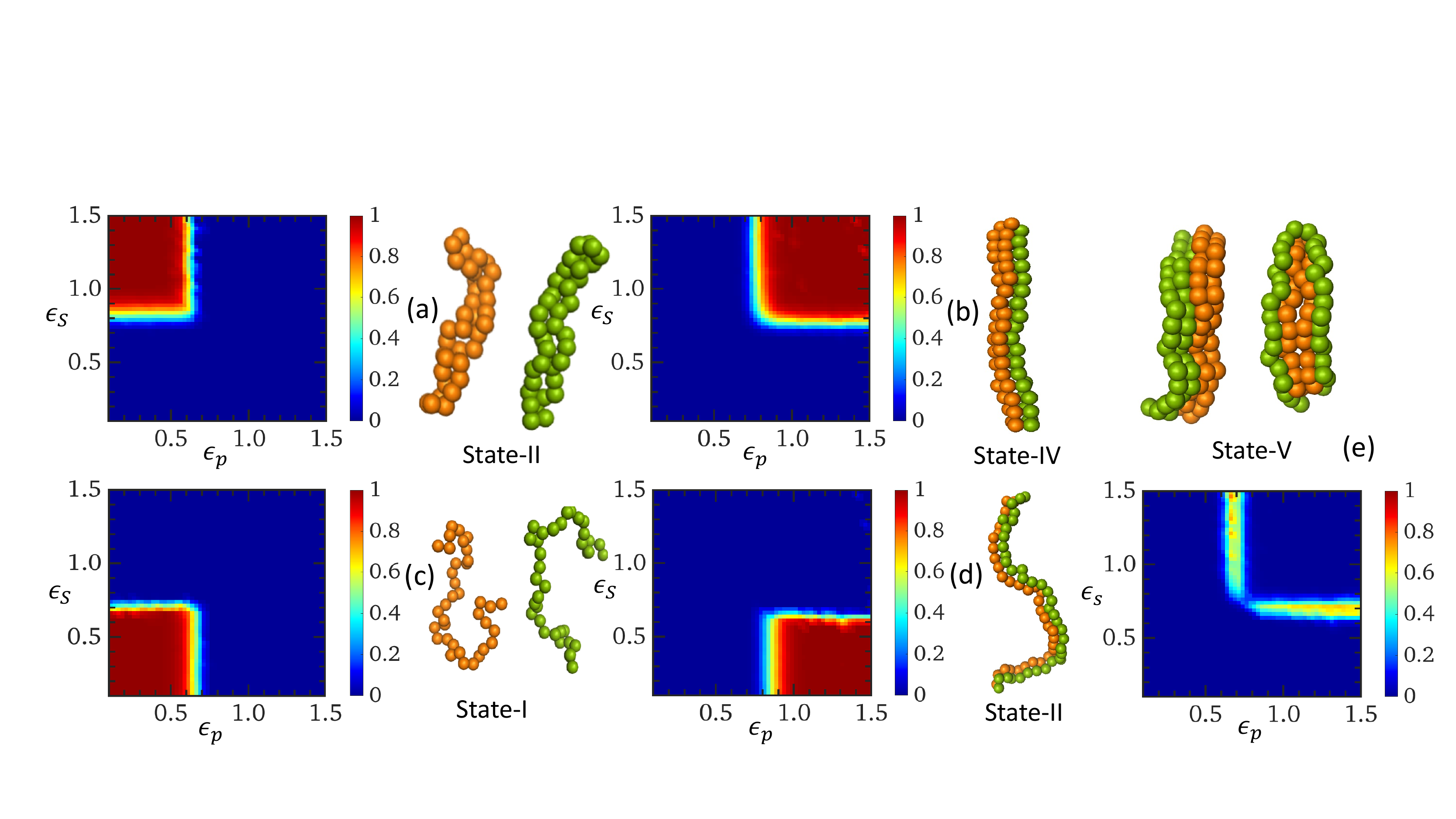}
\caption{The probability of occurrence of different states in the
  phase diagram.  (a-d) the right panel shows four different states of
  polymer A (green) and polymer B (orange); State-I (two free ssDNA),
  State-II (one dsDNA), State-III (two folded ssDNA), and State-IV
  (folded dsDNA). In the left panel, color map shows the probability
  of occurrence of a state (next to it) in different regions of the
  $\epsilon_s $-$ \epsilon_p$ plane. (e) A sheet structure formed from
  polymer A (green) and polymer B (orange) (top panel) and their
  occurrence probability in different regions of the $\epsilon_s$-$
  \epsilon_p$ plane (bottom panel). The sheet structure occurs in the
  Y-regions of the phase diagram.}
\label{fig:prob_state}
\end{figure*}

To address the above question,  The phase boundary is plotted
in Fig. \ref{fig:phase_dia}.  We obtain a horizontal phase boundary
close to $\epsilon_s =0.75$ for $\epsilon_p < 1.00$, which starts
bifurcating into two branches for $\epsilon_p > 1.00$ (wider peak in
energy fluctuations, Fig. S1 \cite{note3}). Similarly, we observe a 
vertical phase boundary close to $\epsilon_p =0.75$ for $\epsilon_s < 0.80$, 
which starts bifurcating into two branches for $\epsilon_s > 0.80$ (Fig. S2 ). 
Overall, the phase diagram is symmetric, except for the point at
which the bifurcation starts appearing in the phase diagram
($\epsilon_p > 1.00$, and $\epsilon_s > 0.80$). In addition to the
horizontal and vertical boundaries, as suggested by the energy
construct, we obtain two additional regions (Y-regions) in the phase
diagram (Fig. \ref{fig:phase_dia}). The average over two boundaries of
each Y-region further corroborates the horizontal or vertical nature
of the boundaries in the phase diagram.

\subsection{Probability of occurrence of 
different phases}

To get the microscopic view of the phase diagram, we probe the
probability of occurrence of different structures in different regions
of the phase diagram. The structures can be quantified by the number
of base pairs ($n_p$) and the number of intra-strand pairs ($n_s$)
present in a given conformation as below (the  maximum number of
allowed pairs is $n_{max} = N$)\\
State-I:  $n_p$ and $n_s$ both are small ($ < 0.3n_{max}$),\\
 State-II:  large $n_p (> 0.8n_{max}$) and small $n_s (< 0.3n_{max})$, \\
 State-III: small $n_p (<0.3n_{max})$ and large $n_s (> 0.8n_{max})$,\hfill \\
State-IV: $n_p$ and $n_s$ both are large ($ > 0.8n_{max}$).\\
The probability of occurrence of different states defined above
($P_{I}$, $P_{II}$, $P_{III}$ and $P_{IV}$) are plotted in Fig.
\ref{fig:prob_state}.  We see from Fig.  \ref{fig:prob_state}c, that
$P_I =1$ for lower values of $\epsilon_p$ and $\epsilon_s$ 
and decreases as we increase $\epsilon_p$ or $\epsilon_s$. For very
large values of $\epsilon_p$ or $\epsilon_s$, $P_I$ becomes zero, and
state II or state III starts appearing with small probabilities (Figs.
\ref{fig:prob_state}d, \ref{fig:prob_state}a).  For larger values of
$\epsilon_p$ and lower values of $\epsilon_s$ (the lower branch of the
horizontal Y-region), $P_{II} =1 $ (Fig. \ref{fig:prob_state}d) while
for larger values $\epsilon_s$ and lower values of $\epsilon_p$ (the
left branch of the vertical Y-region), $P_{III}=1$ (Fig.
\ref{fig:prob_state}a), respectively. These states persist as we
increase $\epsilon_p$ or $\epsilon_s$. State IV starts appearing on
the right of the vertical Y branch and above the horizontal Y branch,
$P_{IV}=1$ (Fig. \ref{fig:prob_state}b) at larger values of
$\epsilon_p$ and $\epsilon_s$, with $P_{IV}=0$ elsewhere.  Many mixed
states (State-VI) appear with finite probability ($P_{VI}$) near the
phase boundaries (coexistence line) (Fig. S3-S4). The observation of
states (I-IV) are in accordance with those predicted by the simple
model of energy considered earlier.  However, the simple model needs
modifications to provide insight about the Y-regions and the
corresponding phases in the phase diagram. 

It should be emphasized here that the vertical (horizontal) Y-region
starts appearing around intermediate values of $\epsilon_p
(\epsilon_s)$, where a large energy $\epsilon_s (\epsilon_p)$ is
already present in the system. A comparable energy scale may stabilize
an intermediate phase between State III (State II) and State IV. As we
cross the vertical Y-region, $n_s$ ($>0.8n_{max}$) is almost unchanged
(for large $\epsilon_s$) and $n_p$ goes from zero ($<0.3n_{max}$ for
small $\epsilon_p$) to maximum ($>0.8n_{max}$ for large $\epsilon_p$).
In the phase diagram, one may look further for a state with
intermediate $n_p$ and its occurrence probability.  Motivated
by this, we define a State-V ($n_s > 0.8 n_{max}$ and $0.35
n_{max}<n_p <0.6 n_{max}$).  This state has a very high occurrence
probability ($P_V \sim 0.6$) as can be seen in Fig. \ref{fig:prob_state}e
color map, and it appears exactly in the vertical Y-region of the phase
diagram though absent elsewhere. This State-V is a sheet-like structure
(Fig.  \ref{fig:prob_state}e, upper left structure), formed as two
folded ssDNA come close to each other.  A similar sheet-like structure
($n_p > 0.8 n_{max}$ and $0.35 n_{max} <n_s <0.6 n_{max}$, Fig.
\ref{fig:prob_state}e, upper right structure) occurs with high
probability in the horizontal Y-region, with the only difference that
$n_s$ and $n_p$ flip their positions. Moreover, two folded ssDNA
(state III, for large $\epsilon_s$) can be close to each other to form
a sheet structure, without any geometrical constraint, by introducing
bonds of comparable energy $\epsilon_p$.  However, forming the sheet
from state II (large $\epsilon_p$) is more difficult energetically.
As the backbone of dsDNA (State-II) consists of springs (first term of
Eq. \ref{eq:1}), the formation of a sheet-like structure requires the
stretching of a few springs at one side of the sheet.  Therefore, the
backbone always puts geometrical constraints on one side of the sheet.
This extra geometric constraint results in the asymmetry in the
starting points of vertical and horizontal Y-regions.

The four states, State-I to State-IV, are polymer-like.  If $s_0$ is
the entropy per monomer of a swollen polymer, we shall have, for the
four states I to IV,
$$S_I=2N s_0, S_{II}=N s_0,\ S_{III}=2 \frac{N}{2} s_0,{\rm \ and\ } S_{IV}=\frac{N}{2}  s_0.$$
The symmetry of the phase diagram follows from the phase boundaries
$$\epsilon_p\big|_{I\rightarrow II} = s_0, {\rm \ and\ } \epsilon_s\big|_{I\rightarrow III} = s_0.$$ 
This picture is validated by more or less vertical or horizontal phase
boundaries (Fig. \ref{fig:phase_dia}), as explained in the previous
paragraph.  In contrast to the four states, State-V takes a flexible
sheet-like structure with an extra source of entropy from the
possibility of folding the sheet.  The free energy of State-V, with
$S_V$ as its entropy, is
$$\frac{F_V}{k_BT}= -\frac{N}{2}\epsilon_p - N \epsilon_s - S_V.$$  
Therefore, the III-V  and the IV-V transitions would take place at
\begin{subequations}
\begin{eqnarray}
    \epsilon_p = 2(S_{III}-S_{V})/N\ {\rm (for\ III\leftrightarrow V)},  \\
    \epsilon_p = 2 (S_V-S_{IV}/N){\ \rm(for \ IV\leftrightarrow V)},
      \label{eq:4}
\end{eqnarray} 
\end{subequations}
which require $S_{III}>S_V>S_{IV}$.  In other words, State-V would
occur in the neighbourhood of the putative III-IV transition.  Similar
arguments can be used for the II-IV phase boundary.

\subsection{Melting of compact DNA}
\begin{figure}[t]
\includegraphics[scale=0.48, trim={3.5in 2.0in 2.85in 1.1in},clip]{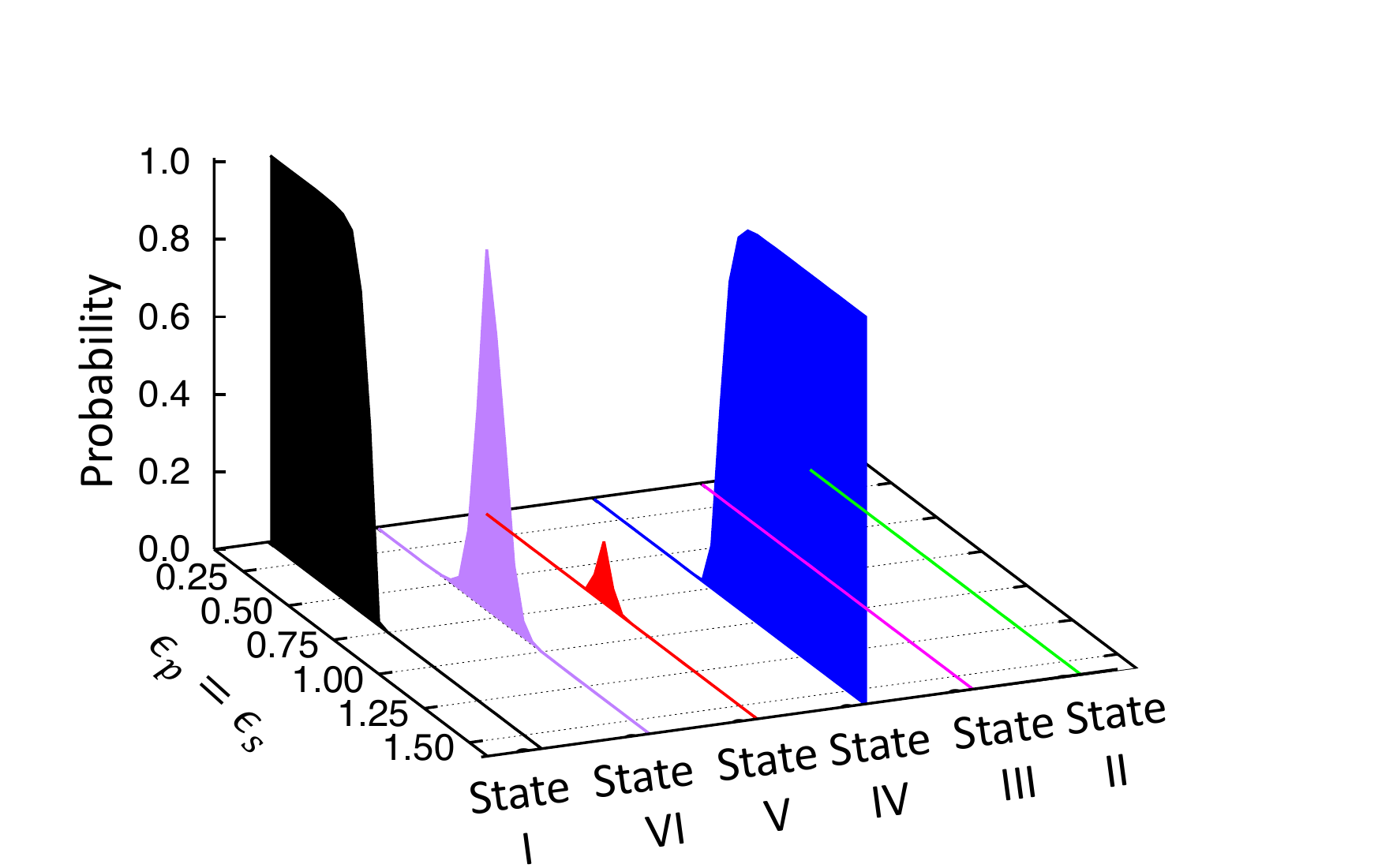}
\caption{Probability of occurrence of different states along the line
  with slope one ($\epsilon_s/\epsilon_p =1 $) in
  Fig.\ref{fig:phase_dia} (dashed line).}
\label{fig:melting}
\end{figure}

The replication and transcription process of DNA requires the
separation of DNA strands.  It is now well established experimentally
and theoretically \cite{smb99,Burg, Prentiss, Koch,
  kumar2010biomolecules, sergei, Kapri, AmitJCP} that DNA undergoes a
temperature (or pH) induced melting transition and a force-induced
unzipping transition, both involving a cooperative breaking of the
base pairs.  
To probe the melting transition of a compact DNA, we follow the evolution 
of different states by varying the interaction energy along the $\epsilon_p/
\epsilon_s =1$ line (the dashed line in Fig.  \ref{fig:phase_dia}).
From our simulations, we compute the occurrence probabilities of
different states (Fig. \ref{fig:melting}).  We start with
$\epsilon_s=\epsilon_p =1.50$, where only state-IV (folded dsDNA)
appears with $P_{IV} =1$ until we are close to the intersection point
in the phase diagram.  Here, $P_{IV}$ decreases, and many mixed
configurations appear which we call state-VI (see Fig. S4,
supplementary material). As the intersection point is a coexistence
point of at least four phases, many mixed configurations are expected
to occur.  In fact, we find $P_{VI} \sim 1$ and $P_V \sim 0.10 $.  The
scenario changes as we go just below the intersection point of the
phase diagram along the chosen line (Fig. \ref{fig:phase_dia}).  $P_{I}$ 
starts picking up and State-I becomes the only state ($P_I =1$) for lower 
values of $\epsilon_s$. In short, we see the melting of a compact DNA along 
the chosen line.  In contrast, for any other line with 
$\epsilon_p / \epsilon_s\neq 1$, the
dsDNA to ssDNA transformation necessarily entails one other state.

\section{Conclusions}

In this paper, we studied the compaction of DNA by considering two
types of effective interactions, (i) inter-strand native-pair
interactions that favour dsDNA formation, and (ii) intra-strand
native-pair interactions that promote the folding of each strand of
DNA.  The varying strengths of these effective interactions mimic the
role of different cellular environments that lead to the compaction
of DNA inside the cell.  We observed five different phases in the
phase diagram as summarized in Fig. \ref{fig:phase_dia}.  There are
transitions taking two free ssDNA (State-I), at lower values of
interaction energies, to either
a bound dsDNA (State-II) or two folded ssDNA (State-III), as one of
the interaction energies (intra-strand or inter-strand) is increased,
keeping the other one at a lower value. For larger values of both
interaction energies, dsDNA folds onto itself (State-IV), a state
reminiscent of a compact DNA.  If a heating process takes a compact
DNA to ssDNA along a line through the intersection point, there will
be a genuine melting point. Otherwise another phase will always
intervene.  Furthermore, we observe from our simulations a sheet-like
arrangement (State-V); it occurs around the region where the III-IV or
the II-IV transitions would have occurred and is unlike any of the
other four states (I to IV), which are polymer-like.  The stability of
the phase comes from the extra folding entropy acquired by a
sheet-like structure.

\section*{Acknowldgments}
GM gratefully acknowledges the financial support from SERB India for a
start-up grant with file Number SRG/2022/001771 and acknowledges the
HPC computing facility at Ashoka University.

%

\begin{thebibliography}{29}
\bibitem {note2} The  words ``compact DNA'' and ``condensed DNA'' will be used interchangeably. Similarly, DNA-condensation and DNA-compaction are used synonymously.
\end{thebibliography}

\pagebreak

\widetext 

\begin{center}
\textbf{\large Supplementary  Material}
\end{center}
\setcounter{equation}{0}
\setcounter{figure}{0}
\setcounter{table}{0}
\setcounter{page}{1}
\setcounter{section}{0}
\renewcommand{\theequation}{S\arabic{equation}}
\renewcommand{\thefigure}{S\arabic{figure}}
\renewcommand{\thesection}{S\arabic{section}}

\begin{center}
{\large{\bf  ``A sheet-like structure in the proximity of compact DNA''}}\\
{Garima Mishra} and {Somendra M. Bhattacharjee}\\
{Department of Physics,  Ashoka University, Sonepat 131029, India}\\

email: garima.mishra@ashoka.edu.in, {somendra.bhattacharjee@ashoka.edu.in}
\end{center}


\appendix

\begin{figure*}[ht]
\includegraphics[scale=0.6, trim={0.0 0in 0.1in 0.in},clip]{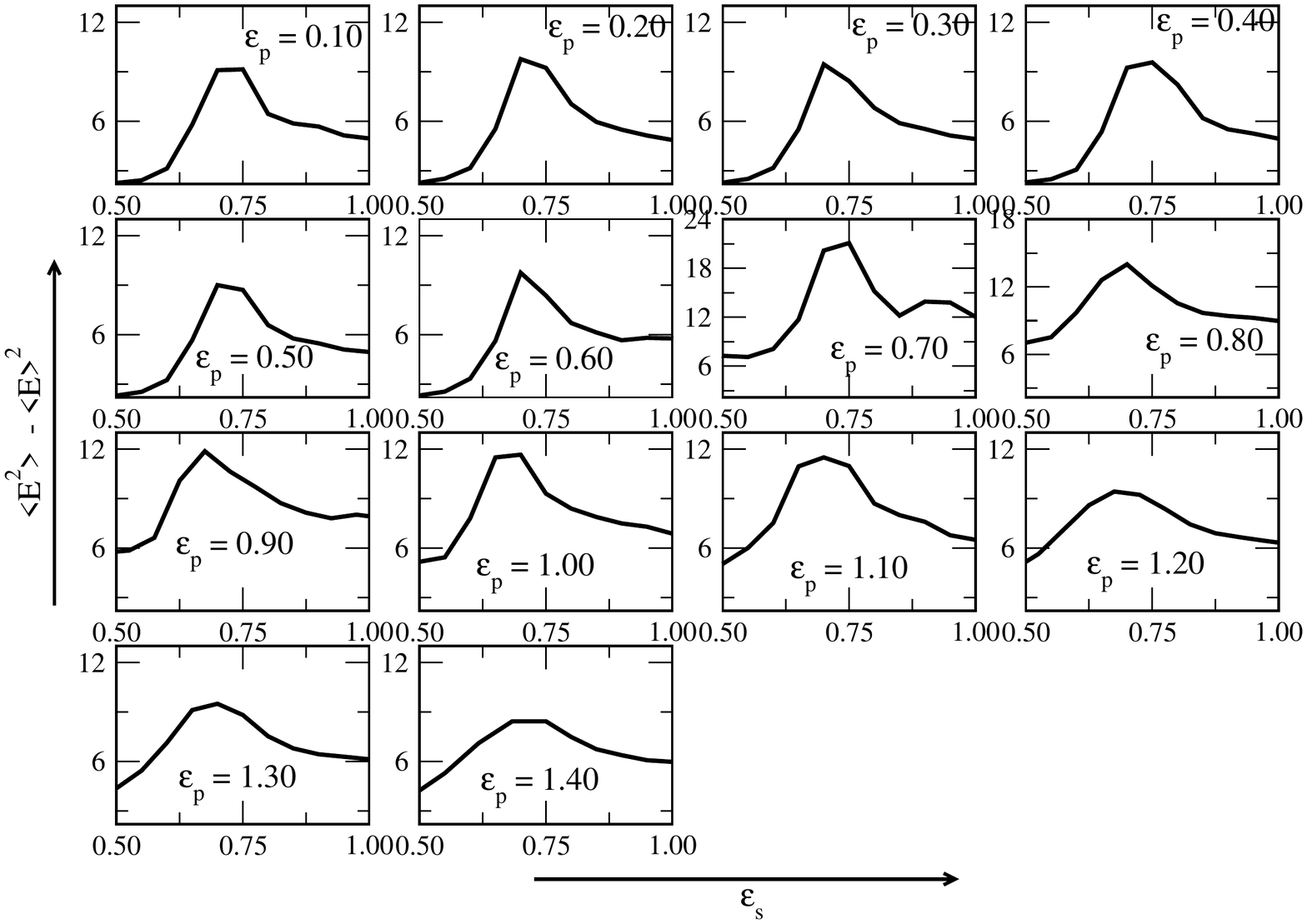}
\caption{\centering{Energy fluctuations as a function of $\epsilon_s$ at different 
values of $\epsilon_p$.}}
\label{fig:Energy_flua}
\end{figure*}

\begin{figure*}[ht]
\includegraphics[scale=0.6, trim={0.0in 0in 0in 0.in},clip]{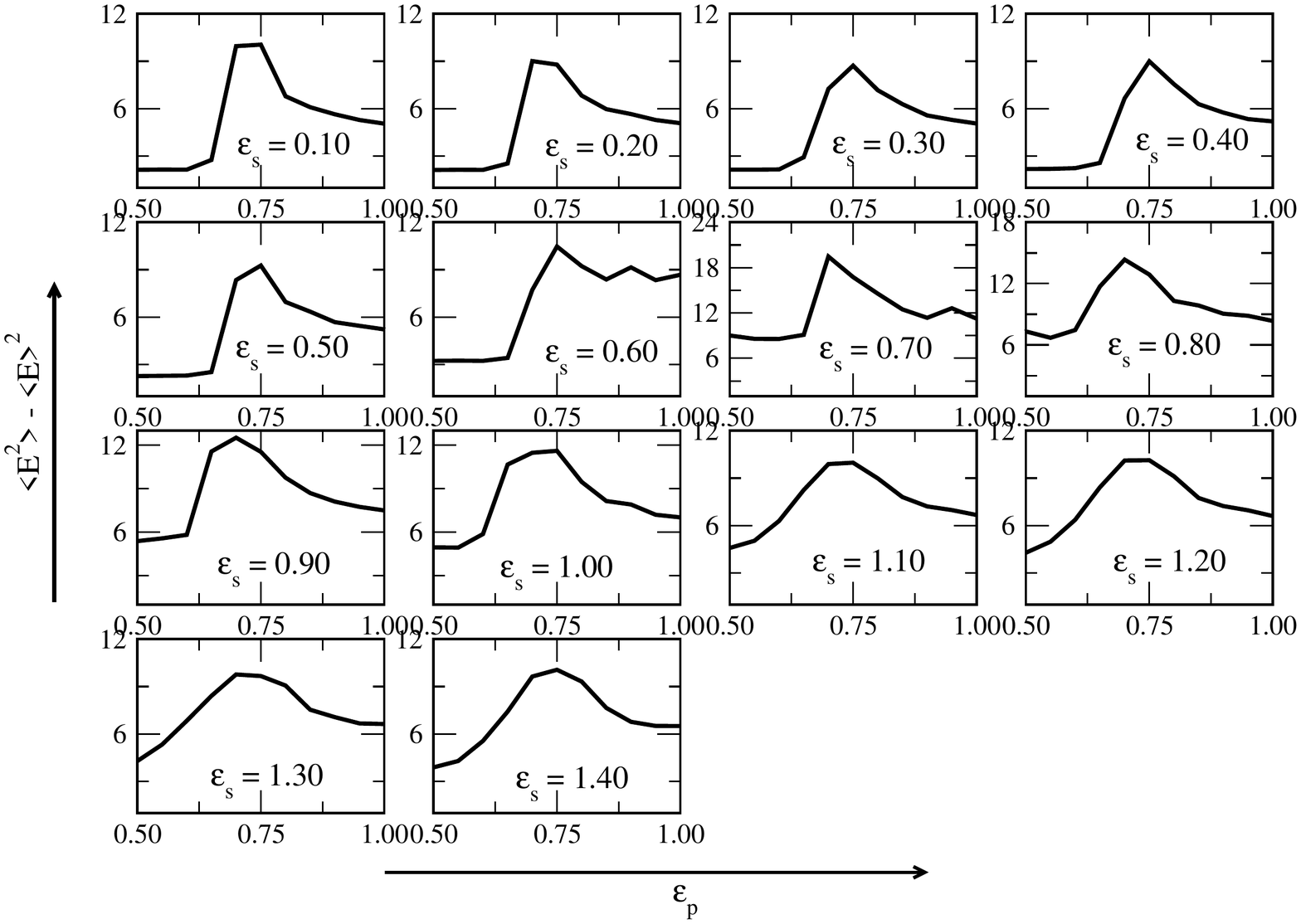}
\caption{Energy fluctuations as a function of $\epsilon_p$ at different values of $\epsilon_s$.}
\label{fig:Energy_flub}
\end{figure*}

\begin{figure*}[htb]
    \centering
    \includegraphics[scale=0.5,trim={0 2.0in 3.5in 0}]{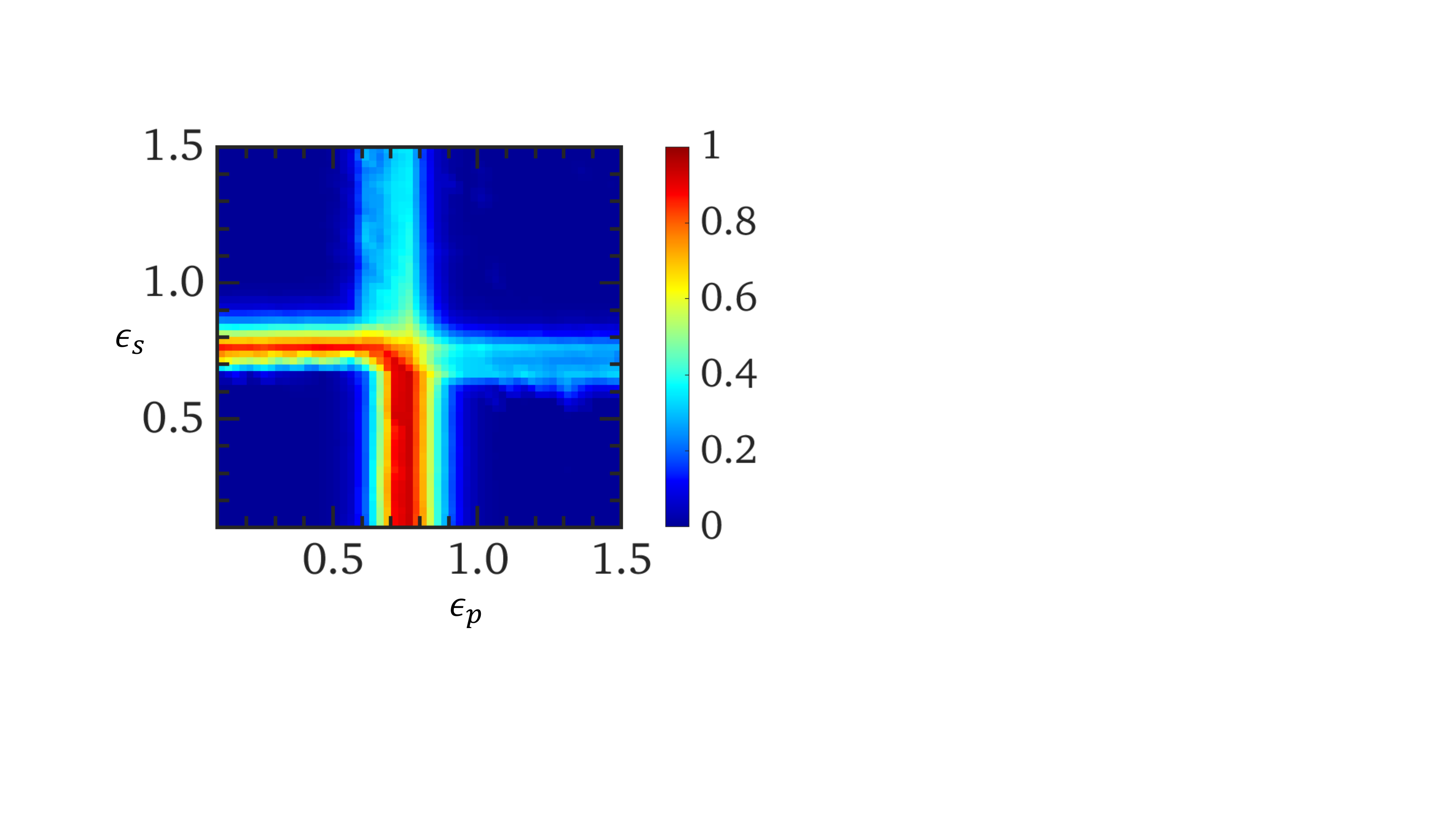}     
    \caption{The probability of occurrence of state-VI (mixed states).}
    \label{fig:prob_VI}
\end{figure*}

\begin{figure*}[ht]
\includegraphics[scale=1.,angle=0,trim={0.76in 0in 6.0in 0.in},clip]{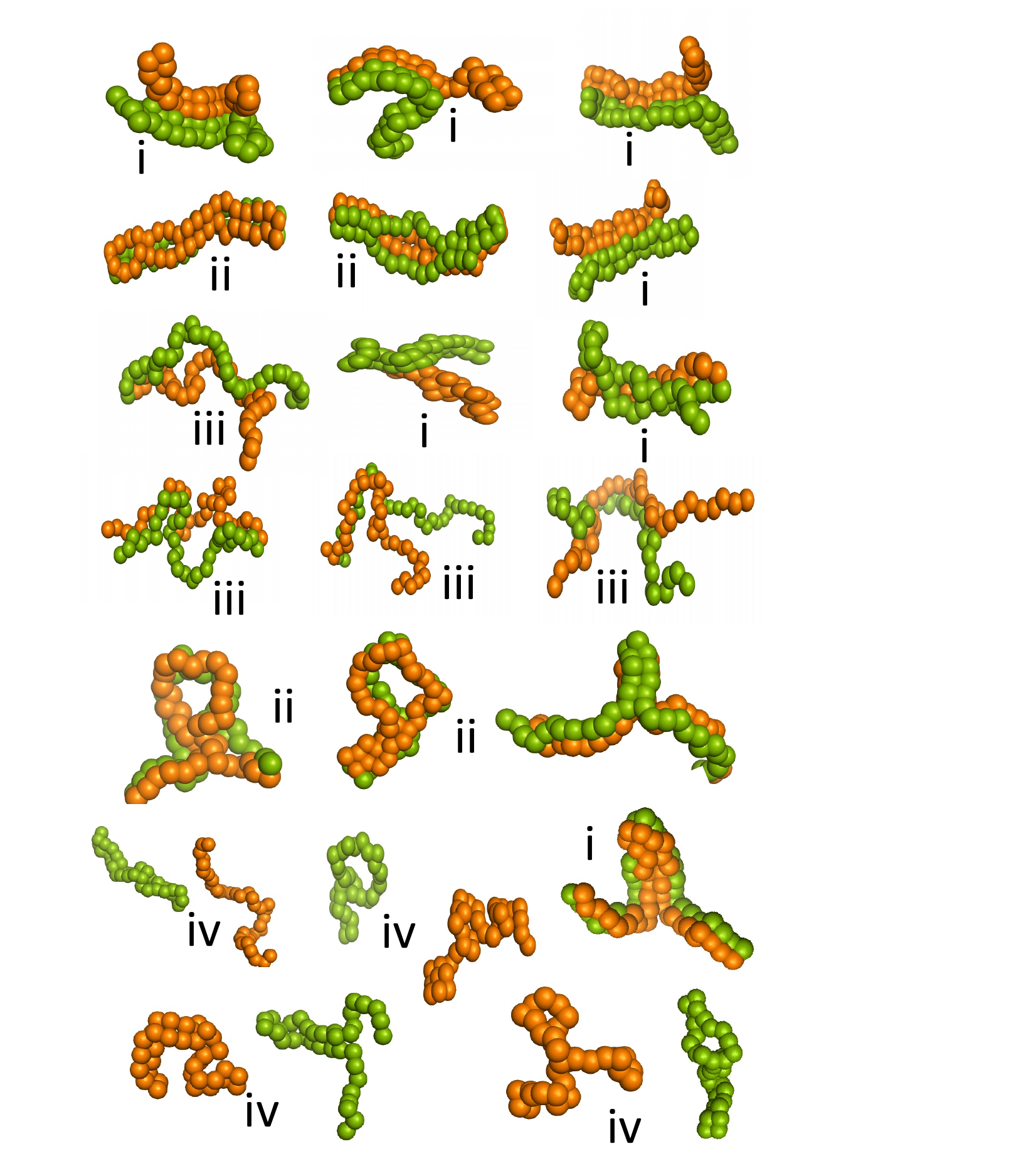}
\caption{Some representative conformations in
state-VI, e.g.  A sheet open at one or both the 
ends(i), folded dsDNA with loop or y-fork(ii), 
A dsDNA with loop or y-fork (iii), folded ssDNA with loop or yfork (iv).  
These are all examples of two phases coexisting along the length of DNA.}
\label{fig:mixed_state_conf}
\end{figure*}

\end{document}